\documentclass[12pt]{article}
\usepackage{amsmath}
\usepackage{verbatim}
\usepackage{latexsym}
\usepackage{graphicx}
\usepackage{axodraw}
\usepackage{multirow}
\usepackage[title]{appendix}

\newcommand{\bqa}{\begin{eqnarray}}
\newcommand{\eqa}{\end{eqnarray}}
\newcommand{\nll}{\nonumber\\}

\newcommand{\sss}[1]{\scriptscriptstyle{#1}} 

\newcommand{\ip }[1]{u\left({#1}        \right)}    
\newcommand{\iap}[1]{{\bar{v}}\left({#1}\right)}    
\newcommand{\sla}[1]{/\!\!\!#1}

\def\mz {M_{\sss{Z}}}
\def\mh {M_{\sss{H}}}
\def\mw {M_{\sss{W}}}
\def\mzs{M_{\sss{Z}}^2}
\def\mhs{M_{\sss{H}}^2}
\def\mws{M_{\sss{W}}^2}

\newcommand{\stw}{s_{\sss{W}}  }
\newcommand{\ctw}{c_{\sss{W}}  }
\newcommand{\cpl}{c_+}
\newcommand{\cmi}{c_-}

\newcommand{\vpael}{\sigma_{e}}
\newcommand{\vmael}{\delta_{e}}
\newcommand{\cosZ}{\cos{\theta_{\sss Z}}}
\newcommand{\aSp}[1]{\left \langle{#1}\right |}
\newcommand{\bSp}[1]{\left [{#1}\right|}
\newcommand{\SpA}[1]{\left |{#1}\right \rangle}
\newcommand{\SpB}[1]{\left |{#1}\right]}
\newcommand{\SpAA}[2]{\left \langle {#1}\:{#2}\right \rangle}

\newcommand{\SpBB}[2]{\left [ {#1}\:{#2}\right ]}
\newcommand{\SpAIA}[2]{\left \langle {#1}|{#2}\right \rangle}

\newcommand{\SpBIB}[2]{\left [ {#1}|{#2}\right ]}

\newcommand{\SpAXB}[3]{\left \langle {#1}|{#2}|{#3}\right ]}
\newcommand{\SpBXA}[3]{\left [ {#1}|{#2}|{#3}\right \rangle}

\newcommand{\SpIAoBI}[2]{\SpA{#1}\!\bSp{#2}}
\newcommand{\SpIBoAI}[2]{\SpB{#1}\!\aSp{#2}}

\newcommand{\labhel}[5]{{}^{\text{#1}}_{{#2}{#3}{#4}{#5}} }

\newcommand{\EL}{e}
 \newcommand{\MZ}{M_Z}

\begin{document}

\title {One-loop electroweak radiative corrections to polarized $e^+e^- \to ZH$}

\author {S. Bondarenko$^{a}$, Ya. Dydyshka$^{b}$, L. Kalinovskaya$^{b}$, \\
L. Rumyantsev$^{b,c}$, R. Sadykov$^{b}$, V. Yermolchyk$^{b}$}

\date{\today}

\maketitle

\vskip 0.3cm

\begin{center}
$^a$ \it Bogoliubov Laboratory of Theoretical Physics, JINR, Dubna, 141980 Russia \\
$^b$ \it Dzhelepov Laboratory of Nuclear Problems, JINR, Dubna, 141980 Russia\\
$^c$ \it Institute of Physics, Southern Federal University, Rostov-on-Don, 344090 Russia
\end{center}

\begin{abstract}
The paper describes high-precision theoretical predictions obtained for the cross sections
of the process $e^+e^- \to ZH$ for future electron-positron
colliders. The calculations performed  using the {\tt SANC} platform taking
into account the full contribution of  one-loop electroweak radiative corrections,
as well as  longitudinal polarization of the initial beams.
Numerical results are given for the energy
range $E_{cm}=250$~GeV --- $1000$~GeV with various polarization degrees.
\end{abstract}

\newpage

\section{Introduction}
The clean signatures of the reactions at $e^+e^-$ colliders (eeC) combined with the effect of polarization of initial particles can
greatly improve the precision of theoretical predictions for various observables of the Standard Model
processes~\cite{MoortgatPick:2005cw}.

The future linear eeC projects such as FCC$_{ee}$, ILC~\cite{homepagesILC} 
CLIC~\cite{homepagesCLIC} are designed to provide polarized beams (up to 80\% for electrons and up to 60\% for positrons). For the
future circular eeC --  CEPC~\cite{homepagesCEPC} and FCC-ee~\cite{homepagesFCCee} the prospects for beam polarizations
are also considered.
Energy range for future eeC will be $250-1000$ GeV, while $250$ GeV is the optimal energy for the Higgs production through the
Higgs-strahlung $e^+ e^- \to  ZH$, which is most important to get the precision measurements of Higgs mass,
spin, CP nature, coupling of Higgs to ZZ and various branching ratios.
Thus, it is important to take  beam polarization into account in theoretical calculations.

At eeC the three main Higgs production processes are the Higgs-strahlung $e^+e^- \to ZH$, the W-fusion
$e^+e^- \to \bar{\nu_e}\nu_e(W^+W^-) \to \bar{\nu_e}\nu_eH$
and the Z-fusion
$e^+e^- \to e^+e^-(ZZ) \to e^+e^-H$
~\cite{Blondel:2018mad,Blondel:2018aan,Gounaris:2014tha,Lafaye:2017kgf,Barger:1993wt,Accomando:1997wt}.

In this paper we present results of the full one-loop electroweak (EW) corrections to the process
\bqa
e^+(p_1) + e^-(p_2)  \rightarrow  Z(p_3) + H(p_4),
\eqa
for arbitrary longitudinal  polarizations $P_{e^+}$ and $P_{e^-}$ of the positron and electron beams, respectively.
Numerical results are evaluated for the following longitudinal polarizations:
(0,0;-0.8,0;-0.8,-0.6;-0.8,0.6)
and for the energies: $250, 500, 1000$ GeV.

The radiative corrections (RC) to $e^+e^- \to ZH$ with unpolarized initial particles were extensively considered in the
literature~\cite{Fleischer:1982af,Kniehl:1991hk,Gong:2016jys}.
The effect of polarization on the virtual and soft photonic contributions to electroweak (EW) RC to Higgs-strahlung process
was  previously calculated in~\cite{Denner:1991ue,Denner:1992bc}.
The present paper also takes into account the hard
Bremsstrahlung contribution.

Numerical estimates are presented for the correction of the total cross section, of the differential distribution in
the Z boson scattering angle~$\cos\vartheta_{Z}$ and for the left-right asymmetry $A_{LR}$ as a function of $\cos\vartheta_{Z}$.
The relevant contributions to the cross section are calculated analytically and then evaluated numerically.

For the numerical evaluation of the process we use the extended version of our Monte Carlo (MC) generator of unweighted events,
that is based on the {\tt SANC}~\cite{Andonov:2008ga} platform, and was previously used for Bhabha process~\cite{Bardin:2017mdd}.
The polarized virtual and soft Bremsstrahlung contributions
are compared with the results of ~\cite{Denner:1992bc}. The cross sections for polarized Born and hard
Bremsstrahlung are cross-checked with the corresponding results of the {\tt WHIZARD}~\cite{Kilian:2018onl}
and {\tt CalcHEP}~\cite{Belyaev:2012qa} programs.

The structure of the paper is the following. In Sect. 2 we describe the cross section calculation technique at the EW one-loop level.
Expressions for covariant (CA) and helicity amplitudes (HA) are presented. The approach to taking into account the
polarization effects is discussed. In Sect.3 we give our numerical results for the total and differential cross sections
and relative corrections. Sect. 4 contains  conclusion and discussion of obtained results.

\section{Differential cross section}
Let us consider scattering of longitudinally polarized $e^+$ and $e^-$ with polarization degrees
$P_{e^+}$ and $P_{e^-}$, respectively. Then the cross section of the generic process $e^+e^- \to ...$ can be expressed as
\begin{equation}
\sigma_{P_{e^-}P_{e^+}} = \frac{1}{4}\sum_{\chi_1,\chi_2}(1+\chi_1P_{e^-})(1+\chi_2P_{e^+})\sigma_{\chi_1\chi_2},
\label{eq1}
\end{equation}
where $\chi_i = -1(+1)$ corresponds to lepton with left (right) helicity state. Thus the cross section with
arbitrary longitudinal polarization is a linear combination of four contributions:
\bqa
\sigma_{--(-+,+-,++)} \equiv \sigma_{LL,(LR,RL,RR)}.
\eqa

At one-loop the cross section of the process can be divided into four parts:
\bqa
\sigma^{\text{1-loop}} = \sigma^{\mathrm{Born}} + \sigma^{\mathrm{virt}}(\lambda)
+ \sigma^{\mathrm{soft}}(\lambda,\omega) + \sigma^{\mathrm{hard}}(\omega),
\eqa
where $\sigma^{\mathrm{Born}}$ --- Born level cross-section,
$\sigma^{\mathrm{virt}}$ --- contribution of virtual(loop) corrections,
$\sigma^{\mathrm{soft}}$ --- contribution due to soft photon emission,
$\sigma^{\mathrm{hard}}$ --- contribution due to hard photon emission
(with energy { $E_{\gamma} > \omega$}). Auxiliary parameters $\lambda$ ("photon mass")
and $\omega$ are cancelled out after summation.

We count all contributions through helicity amplitudes approach.

The virt(Born) cross section of the $e^+e^- \to ZH$ process can be written as:
\bqa
\frac{d\sigma\labhel{virt(Born)}{ \chi_1}{ \chi_2}{}{}}{d\cosZ}
= \frac{\sqrt{\lambda(s,\mzs,\mhs)}}{32\pi s^2}|\mathcal{H}\labhel{virt(Born)}{ \chi_1}{ \chi_2}{}{}|^2,
\eqa
where
\bqa
|\mathcal{H}\labhel{virt(Born)}{ \chi_1}{ \chi_2}{}{}|^2 = \sum_{\chi_3 = 0,\pm1} |\mathcal{H}\labhel{virt(Born)}{ \chi_1}{ \chi_2}{\chi_3}{}|^2.
\eqa

The soft term is factorized to Born-level cross section:
\bqa
\frac{d\sigma\labhel{soft}{ \chi_1}{ \chi_2}{}{} }{d\cosZ}=
\frac{d\sigma\labhel{Born}{ \chi_1}{ \chi_2}{}{} }{d\cosZ}
\cdot \frac{\alpha}{2\pi}
\left(- L_s^2 + 4L_s\ln \frac{2\omega}{\lambda}
 - \frac{2\pi^2}{3}  + 1\right)\!\!, \quad L_s=\ln\frac{s}{m_e^2}-1.
\eqa

The cross section for hard Bremsstrahlung $ e^+(p_1) + e^-(p_2)  \rightarrow  Z(p_3) + H(p_4) + \gamma(p_5) $ is
\bqa
\frac{d\sigma\labhel{hard}{ \chi_1}{ \chi_2}{}{} }{ds'
	d\cos{\theta_4 }d\phi_4
	d\cos{\theta_5}}
= \frac{s-s'}{8(4\pi)^4ss'}\frac{\sqrt{\lambda(s',\mzs,\mhs)}}{\sqrt{\lambda(s,m_e^2,m_e^2)}}
|\mathcal{H}\labhel{hard}{ \chi_1}{ \chi_2}{}{} |^2,
\eqa
where $s'=(p_3+p_4)^2$, and 
\bqa
|\mathcal{H}\labhel{hard}{ \chi_1}{ \chi_2}{}{} |^2 = \sum_{\chi_3 = 0,\pm1}\sum_{\chi_5 = \pm1} |\mathcal{H}\labhel{hard}{ \chi_1}{ \chi_2}{ \chi_3}{\chi_5} |^2.
\eqa

Here $\theta_5$ is an angle between 3-momenta of the photon and the positron,
$\theta_4$ --- an angle between 3-momenta of the $Z$-boson and the photon in the rest
frame of
$(Z,H)$-compound, $\phi_4$ --- an azimuthal angle of $Z$-boson in the rest frame of
$(Z,H)$-compound.

\subsection{Covariant amplitude for virtual parts and Born level}

The covariant amplitude neglecting the masses
of initial particles can be written as~\cite{Bardin:2005dp}:
\bqa
{\cal A}^{\sss {eeZH}} &=& N(s) \,\Biggl\{
\Biggl[\iap{p_1} \biggl(
 \gamma_{\nu}\gamma_{+}  \vpael  {\cal F}^{+}_0(s,t)
  +\sum_{i=1,2}\sla{p_3} \gamma_{+}  (p_i)_{\nu} {\cal F}^{+}_i(s,t)\biggr)
  \ip{p_2} \varepsilon^{\sss Z}_{\nu}(p_3)
\Biggr]
\nll &&
\phantom{-\frac{i g^2}{8 \ctw^2} }
+ \Biggl[\vpael \to \vmael,\;\gamma_{+}\to\gamma_{-},\;
{\cal F}^{+}_i(s,t)\to {\cal F}^{-}_{i}(s,t)
  \Biggr]\Biggr\},
\label{ffHZ-ann}
\eqa
\noindent where
\vspace*{-5mm}
\bqa
N(s) = \frac{i g^2}{4 \ctw^2} \frac{\mz}{s-\mz^2+i \mz \Gamma_{\sss Z}}.
\eqa
We also use various coupling constants
\bqa
\sigma_e = v_e + a_e\,,\quad \delta_e = v_e - a_e\,,
\quad \ctw=\frac{\mw}{\mz}\,,\quad
g=\frac{e}{\stw}\,,\quad \mbox{\it etc.}
\eqa

\subsection{Helicity amplitudes for virtual parts and Born level}

$\bullet$ {HA for virtual part}

There are 6 non-zero HAs for virtual contribution:
\bqa
{\cal H}_{+-+} &=&  N(s)  \sqrt{\dfrac{  s}{  2}} \cpl
 \Bigl\{ \sqrt{\lambda} c_- \left[{\cal F}^{+}_2(s,t)-{\cal F}^{+}_1(s,t)\right] - 4 \vpael {\cal F}^{+}_0(s,t)\Bigr\},\\
%
 {\cal H}_{+--} &=&  N(s) \sqrt{\dfrac{  s}{  2}} \cmi
 \Bigl\{ \sqrt{\lambda} c_+ \left[{\cal F}^{+}_2(s,t)-{\cal F}^{+}_1(s,t)\right]  - 4\vpael {\cal F}^{+}_0(s,t)\Bigr\},\nll
 {\cal H}_{+-0} &=&  N(s) \dfrac{  \sin \vartheta_z}{  2 \mz } 
 \Bigl\{\sqrt{\lambda}
 \left[ \beta_+ {\cal F}^{+}_1(s,t)
       +\beta_- {\cal F}^{+}_2(s,t)\right]
       + 4 \vpael L {\cal F}^{+}_0(s,t)
   \Bigr\},\nonumber
\eqa
 where
\bqa
& L={s+\mz^2-\mh^2},\quad \lambda=\lambda(s,\mz^2,\mh^2)\,, \\
&\beta=\beta(s,\mz^2,\mh^2)=\dfrac{ \sqrt{\lambda}}{ L},\quad
 \beta_{\pm}=\beta \pm \cos{\vartheta_{\sss{Z}}}, \quad c_\pm = 1 \pm \cos{\vartheta_{\sss{Z}}}.
\eqa

Expression for the amplitude ${\cal H}_{-++}$ is obtained from the expression ${\cal H}_{+-+}$ by the replacement
$(\vpael \to \vmael, c_+ \to c_-, {\cal F}^{+} \to {\cal F}^{-} )$, the same procedure is applied
to obtain ${\cal H}_{-+-}$ from ${\cal H}_{+--}$. To obtain amplitude ${\cal H}_{-+0}$ from $-{\cal H}_{+-0}$ 
the replacement  $(\vpael \to \vmael)$ works. 

$\bullet$ {HA for Born level}

In order to get helicity amplitudes for the Born level one should set ${\cal F}^{\pm}_i(s,t)=0$ and ${\cal F}^{\pm}_0(s,t)=1$.

\clearpage

\subsection{Helicity amplitudes for hard Bremsstrahlung}

For massless particle with light-like momentum $k_i$ we are using following notations for spinors:
		\begin{gather} 	
\begin{aligned}  
u_{+}(k_i) &=\gamma_{+} u(k_i) = v_{-}(k_i) = \gamma_{+} v(k_i) =\SpA{i},
\\ 
u_{-}(k_i) &=\gamma_{-} u(k_i) = v_{+}(k_i) = \gamma_{-} v(k_i) =\SpB{i},
\\
\bar{u}_{+}(k_i) &= \bar{u}(k_i)\gamma_{-} = \bar{v}_{-}(k_i) =  \bar{v}(k_i)\gamma_{-} =\bSp{i},
\\
\bar{u}_{-}(k_i) &= \bar{u}(k_i)\gamma_{+} = \bar{v}_{+}(k_i) =  \bar{v}(k_i)\gamma_{+} =\aSp{i}.
\end{aligned}
\end{gather}
Using them we able to construct polarization wave-functions for other particles, including massive.

We project all massive momenta with $p_i^2=m_i^2$ to the light-cone of photon $p_5$ and introduce associated ``momenta'':
\vspace{-7pt}
\begin{gather}
\begin{aligned}  
k_i &= p_i - \dfrac{m_i^2}{2p_i\cdot p_5}p_5, & k_i^2&=0,  & i &=1..4,
\end{aligned}	
\\
\begin{aligned}  
k_5 &= -\sum_{i=1}^{4}k_i = K p_5, &  K &= 1+ \sum_{i=1}^4\dfrac{m_i^2}{2p_i\cdot p_5}= 1+ \sum_{i=1}^4\dfrac{m_i^2}{2k_i\cdot p_5},
\\
p_5 &= -\sum_{i=1}^{4}p_i = K' k_5, &  K' &= 1- \sum_{i=1}^4\dfrac{m_i^2}{2p_i\cdot k_5}  = 1- \sum_{i=1}^4\dfrac{m_i^2}{2k_i\cdot k_5} .
\end{aligned}	
\end{gather}
Vector $k_5$ appears to be light-like, so we are left with ``momentum conservation'' of associated vectors.

Construction of photon polarization vector needs introduction auxilary massless vector $q$ for gauge fixing. Physical amplitudes are independent of it. We use following parametrization:
\begin{gather*}
\begin{aligned}
\varepsilon^{+}_{\mu}(k_5) &=\dfrac{\SpAXB{q}{\gamma_\mu}{5}}{\sqrt{2}\SpAIA{q}{5}},
&
\varepsilon^{-}_\mu(k_5) &=\dfrac{\SpBXA{q}{\gamma_\mu}{5}}{\sqrt{2}\SpBIB{q}{5}}
= (\varepsilon^{+}_\mu(k_5))^{*},
\end{aligned}
\\
\begin{aligned}
  \hat{\varepsilon}^{+}(k_5)=\gamma^\mu \varepsilon^{+}_{\mu}(k_5) &=
         \sqrt{2}\dfrac{  \SpIAoBI{q}{5}+\SpIBoAI{5}{q}}{  \SpAIA{q}{5}}, & \hat{\varepsilon}^{-}(k_5) &=\sqrt{2}\dfrac{  \SpIBoAI{q}{5}+\SpIAoBI{5}{q}}{  \SpBIB{q}{5}}.
\end{aligned}
\end{gather*}  

There are 20 non-zero HAs for hard contribution:

\begin{gather*}
\begin{aligned}
A\labhel{}{-}{-}{+}{+} &= 2\EL m_{1} \MZ  N(s') \bigg( \dfrac{ \vmael}{s_{15}} +   \dfrac{\vpael}{s_{25}}\bigg)  \SpBIB{1}{2}\dfrac{ \SpAIA{3}{5}}{\SpBIB{3}{5}},
\\
A\labhel{}{+}{+}{-}{+} &=2\EL m_{1} \MZ N(s') \bigg(  \dfrac{  \vpael}{s_{15}} +   \dfrac{ \vmael}{s_{25}}\bigg)\SpBIB{1}{2}\dfrac{ \SpBIB{3}{5} \SpAIA{1}{5}\SpAIA{2}{5}}{\SpAIA{3}{5}\SpBIB{1}{5}\SpBIB{2}{5}},
\\
A\labhel{}{-}{+}{-}{+} &= -2 \EL \MZ N(s')  \dfrac{ \vpael}{s_{15}} \dfrac{\SpBIB{1}{2} \SpBIB{1}{3}\SpAIA{1}{5}\SpAIA{2}{5}}{  \SpAIA{3}{5}\SpBIB{2}{5} },
\\
A\labhel{}{+}{-}{-}{+} &= -2 \EL \MZ N(s')  \dfrac{ \vmael}{s_{25}} \dfrac{\SpBIB{1}{2} \SpBIB{2}{3}\SpAIA{2}{5}\SpAIA{1}{5}}{  \SpAIA{3}{5}\SpBIB{1}{5} },
\\
A\labhel{}{-}{-}{0}{+} &= \sqrt{2}\EL m_{1}   N(s') \bigg(\dfrac{ \vmael}{s_{15}}\dfrac{  \SpBIB{2}{3}}{\SpBIB{2}{5}} + \dfrac{ \vpael}{s_{25}} \dfrac{\SpBIB{1}{3}}{\SpBIB{1}{5}}\bigg) \SpBIB{1}{2}\SpAIA{3}{5},
\end{aligned}
\end{gather*}

\begin{gather*}
\\
\begin{aligned}
A\labhel{}{-}{+}{+}{+} &= -2 \EL \MZ  N(s')  \vpael \bigg(\dfrac{\SpBIB{1}{2}\SpAIA{2}{3}\SpAIA{2}{5}}{s_{25}\SpBIB{3}{5}}+\dfrac{\SpBIB{1}{5}\SpAIA{3}{5}}{\SpBIB{2}{5}\SpBIB{3}{5}}\bigg),
\end{aligned}
\\
\begin{aligned}
A\labhel{}{+}{-}{+}{+} &= -2 \EL \MZ  N(s') \vmael \bigg(\dfrac{\SpBIB{1}{2}\SpAIA{1}{3}\SpAIA{1}{5}}{s_{15}\SpBIB{3}{5}}-\dfrac{\SpBIB{2}{5}\SpAIA{3}{5}}{\SpBIB{1}{5}\SpBIB{3}{5}}\bigg),
\end{aligned}
 \end{gather*}

\begin{gather*} 
\\
\begin{aligned}
A\labhel{}{+}{+}{0}{+} &=\sqrt{2}  \EL m_{1} N(s') \bigg(\SpBIB{1}{2}\Big(\dfrac{\vpael}{s_{15}}\SpAIA{1}{5}\SpAIA{2}{3}+\dfrac{\vmael}{s_{25}}\SpAIA{2}{5}\SpAIA{1}{3}\Big)+\SpAIA{3}{5}\Big(\vpael-\vmael\Big) \bigg)\dfrac{\SpBIB{3}{5}}{\SpBIB{1}{5}\SpBIB{2}{5}},
\end{aligned} 
\\
\begin{aligned}
A\labhel{}{-}{+}{0}{+} = -\sqrt{2}\EL N(s') &\bigg(
\vpael \Big(\dfrac{\SpBIB{1}{3}\SpAIA{2}{3}\SpAIA{1}{5}}{s_{15}} + \dfrac{\SpBIB{1}{3}\SpAIA{3}{5}}{\SpBIB{1}{2}} +\dfrac{M^2_Z\SpAIA{2}{5}}{s_{45}} \Big)
+ \vmael\dfrac{ m_{1}^{2} s_{45} \SpAIA{2}{5}}{s_{15}s_{25}}\bigg)\dfrac{\SpBIB{1}{2}}{\SpBIB{2}{5}},
\end{aligned}
\\
\begin{aligned}
A\labhel{}{+}{-}{0}{+} = -\sqrt{2}\EL  N(s')  & \bigg(
\vmael \Big(\dfrac{\SpBIB{2}{3}\SpAIA{1}{3}\SpAIA{2}{5}}{s_{25}} - \dfrac{\SpBIB{2}{3}\SpAIA{3}{5}}{\SpBIB{1}{2}} +\dfrac{M^2_Z\SpAIA{1}{5}}{s_{45}} \Big)
+ \vpael\dfrac{ m_{1}^{2} s_{45} \SpAIA{1}{5}}{s_{15}s_{25}}\bigg)\dfrac{\SpBIB{1}{2}}{\SpBIB{1}{5}},
\end{aligned}
\end{gather*}
 where
\bqa
& s_{i5}=2k_i\cdot p_5=K'  \SpAIA{i}{5} \SpBIB{5}{i}.
\eqa

Other ones can be obtained using  CP-symmetry:
\begin{gather*}
\begin{aligned}
{A}\labhel{\text{hard}}{\chi_1}{\chi_2}{\chi_3}{\chi_4} =
-&{\chi_1}{\chi_2}{\chi_3}{\chi_5}
\overline{ {A} }\labhel{\text{hard}}{-\chi_1}{-\chi_2}{-\chi_3}{-\chi_5}|_{p_1\leftrightarrow p_2}. \\
\end{aligned}
\end{gather*}

Freedom in the light-cone projection choice corresponds to arbitrariness of spin quantization direction. We exploit it to make expressions compact. 

To obtain amplitudes $\mathcal{H}$ with definite helicity state spin-rotation matrices should be applied  for each index $\chi$ of incoming particles independently:
\begin{gather*}
\begin{aligned}
\mathcal{H}_{... \xi_i ...} &= C_{\xi_i}^{\phantom{a}\chi_i}\mathcal{A}_{... \chi_i ...}
\end{aligned}
\\
\begin{aligned}
C_{\xi_i}^{\phantom{a}\chi_i}  &= \left[\begin{matrix}
\dfrac{\SpBB{k_{i^\flat}\!}{p_5}}{\SpBB{p_i}{p_5}}
&
\dfrac{m_i\SpAA{k_{i^*}}{p_5}}{\SpAA{k_{i^*}}{k_{i^\flat}}\SpAA{p_i}{p_5}}
\\
\dfrac{m_i\SpBB{k_{i^*}}{p_5}}{\SpBB{k_{i^*}}{k_{i^\flat}}\SpBB{p_i}{p_5}}
&
\dfrac{\SpAA{k_{i^\flat}}{p_5}}{\SpAA{p_i}{p_5}}
\end{matrix}\right]
= 
\left[\begin{matrix}
\dfrac{\SpAA{k_{i^*}}{p_i}}{\SpAA{k_{i^*}}{k_{i^\flat}}}
&
\dfrac{m_i\SpAA{k_{i^*}}{p_5}}{\SpAA{k_{i^*}}{k_{i^\flat}}\SpAA{p_i}{p_5}}
\\
\dfrac{m_i\SpBB{k_{i^*}}{p_5}}{\SpBB{k_{i^*}}{k_{i^\flat}}\SpBB{p_i}{p_5}}
&
\dfrac{\SpBB{k_{i^*}}{p_i}}{\SpBB{k_{i^*}}{k_{i^\flat}}}
\end{matrix}\right]\!\!,
\end{aligned}
\end{gather*}
where
\begin{gather*}
\begin{aligned}
k_{i^*} &= \{|\vec{p}_i|,-p_i^x,-p_i^y,-p_i^z\}, & k_{i^*}^2 &=0,
\\
k_{i^\flat} &= p_i-\dfrac{m_i^2}{2p_i\cdot k_{i^*}}k_{i^*}, & k_{i^\flat}^2 &=0.
\end{aligned}
\end{gather*}

\clearpage

\section{Numerical results and comparison}

In this section, we present the numerical results for the EW RC to $e^+e^- \to HZ$
obtained with help of {\tt SANC}.
We work in $\alpha(0)$-sheme and use the following set of input parameters:
\begin{eqnarray}
&&\alpha^{-1}(0) = 137.03599976, \quad \Gamma_Z = 2.49977 \; \text{GeV}\nonumber\\
&&\mw = 80.4514958 \; \text{GeV}, \quad \mz = 91.1876 \; \text{GeV},
\quad \mh = 125 \; \text{GeV}, \nonumber\\
&&m_e = 0.51099907 \; \text{MeV}, \quad m_\mu = 0.105658389 \; \text{GeV},
\quad m_\tau = 1.77705 \; \text{GeV}, \nonumber\\
&&m_d = 0.083 \; \text{GeV}, \quad m_s = 0.215 \; \text{GeV},
\quad m_b = 4.7 \; \text{GeV}, \nonumber\\
&&m_u = 0.062 \; \text{GeV}, \quad m_c = 1.5 \; \text{GeV},
\quad m_t = 173.8 \; \text{GeV}.
\end{eqnarray}

For the real photon emission we apply the cut on the photon energy $E_{\gamma} > 1$ GeV.

In~\cite{Bardin:2005dp} we compared the results for one-loop EW corrections (excluding hard Bremsstrahlung)
with the results of~\cite{Denner:1992bc} and the program {\tt Grace-Loop}~\cite{Belanger:2003sd}.

In this paper in order to cross-check the results of hard and Born cross sections  we produce the results for these contributions
with the help of the {\tt WHIZARD} and {\tt CalcHEP} programs.
We receive complete agreement in all digits.

$\bullet$ {Energy dependence}

Tables~\ref{Table:results1} -- \ref{Table:results3} show our results
for polarized Born, hard Bremsstrahlung and 1-loop cross sections
and relative correction $\delta$ in percents, which is defined as
\bqa
\delta = \frac{\sigma^{\text{1-loop}}-\sigma^{\text{Born}}}{\sigma^{\text{Born}}} \cdot 100 \%,
\eqa
for various energies and polarization degrees of initial particles.

\begin{table}[!h]
\begin{center}
\begin{tabular}{|r|l|l|l|l|l|}
\hline
$P_{e^-}$ & $P_{e^+}$       &$\sigma^{\text{hard}}$, fb & $\sigma^{\text{Born}}$, fb & $\sigma^{\text{1-loop}}$, fb & $\delta$, \%\\
\hline
\hline
  0      & \phantom{-}0   &  82.0(1)               & 225.59(1)                & 206.91(1)                  & -8.28(1)   \\
 -0.8    & \phantom{-}0   &  47.6(1)               & 266.05(1)                & 223.52(2)                  & -15.99(1)  \\
 -0.8    &           -0.6 &  46.3(1)               & 127.42(1)                & 111.76(2)                  & -12.29(1)\\
 -0.8    & \phantom{-}0.6 & 147.1(1)               & 404.69(1)                & 335.28(1)                  & -17.15(1)\\
\hline
\end{tabular}
\caption{Hard, Born and 1-loop cross sections in fb  of the process $e^+e^- \to ZH$ and relative correction
$\delta$ in percents for energy $250$ GeV and various polarizations of initial particles produced by {\tt SANC}.}
\label{Table:results1}
\end{center}
\end{table}
As it can be seen in Table \ref{Table:results1}, taking into account the polarization significantly affects the value of
the observed cross section: at zero beam polarization the correction value is negative and equal to $-8.28(1)\%$,
and with different set of the  polarization beams the correction remains negative and is two time bigger,
up to $-17.15\%$.

\begin{table}[!h]
\begin{center}
\begin{tabular}{|r|l|l|l|l|l|}
\hline
$P_{e^-}$ & $P_{e^+}$ &  $\sigma^{\text{hard}}$, fb & $\sigma^{\text{Born}}$, fb & $\sigma^{\text{1-loop}}$, fb & $\delta$, \%\\
\hline
\hline
  0      &   \phantom{-}0     &    38.95(1) & 53.74(1) & 62.43(1)  & 16.17(1)\\
 -0.8    &   \phantom{-}0     &    45.92(1) & 63.38(1) & 68.32(1)  & 7.80(1)\\
 -0.8    &   -0.6             &    22.10(1) & 30.35(1) & 34.04(1)  & 12.16(1) \\
 -0.8    &   \phantom{-}0.6   &    69.74(1) & 96.40(1) & 102.60(1) & 6.43(1)\\
\hline
\end{tabular}
\caption{Hard, Born and 1-loop cross sections in fb of the process $e^+e^- \to ZH$ and relative correction
$\delta$ in percents for energy $500$ GeV and various polarizations of initial particles produced by {\tt SANC}.}
\label{Table:results2}
\end{center}
\end{table}
\begin{table}[!h]
\begin{center}
\begin{tabular}{|r|l|l|l|l|l|}
\hline
$P_{e^-}$ & $P_{e^+}$  &$\sigma^{\text{hard}}$, fb & $\sigma^{\text{Born}}$, fb & $\sigma^{\text{1-loop}}$, fb & $\delta$, \%\\
\hline
\hline
  0      &\phantom{-}0     &  11.67(1) & 12.05(1)  & 14.58(1)  & 20.97(1)\\
 -0.8    &\phantom{-}0     &  13.75(1) & 14.217(1) & 15.824(1) & 11.31(1)\\
 -0.8    &          -0.6   &  6.65(1)  & 6.809(1)  & 7.955(1)  & 16.84(1)\\
 -0.8    &\phantom{-}0.6   &  20.85(1) & 21.62(1)  & 23.69(1)  & 9.57(1) \\
 \hline
\end{tabular}
\caption{Hard, Born and 1-loop cross sections in fb  of the process $e^+e^- \to ZH$ and relative correction
$\delta$ in percents  for energy $1000$ GeV and various polarizations of initial particles produced by {\tt SANC}.}
\label{Table:results3}
\end{center}
\end{table}

From  Tables \ref{Table:results2} and \ref{Table:results3}, we see
that at zero beam polarization the correction value is positive and equal $16.17(1)$ for $500$ GeV and
$20.97(1)$ for $1000$ GeV,
and with different set of the  polarization beams the correction remains positive  and varies greatly, up to  $6.43\%$ for $500$ GeV and
$9.57$ for $1000$ GeV.

\newpage

$\bullet$ {Angular dependence}

\begin{figure*}[!h]
\begin{center}
\includegraphics[width=\textwidth]{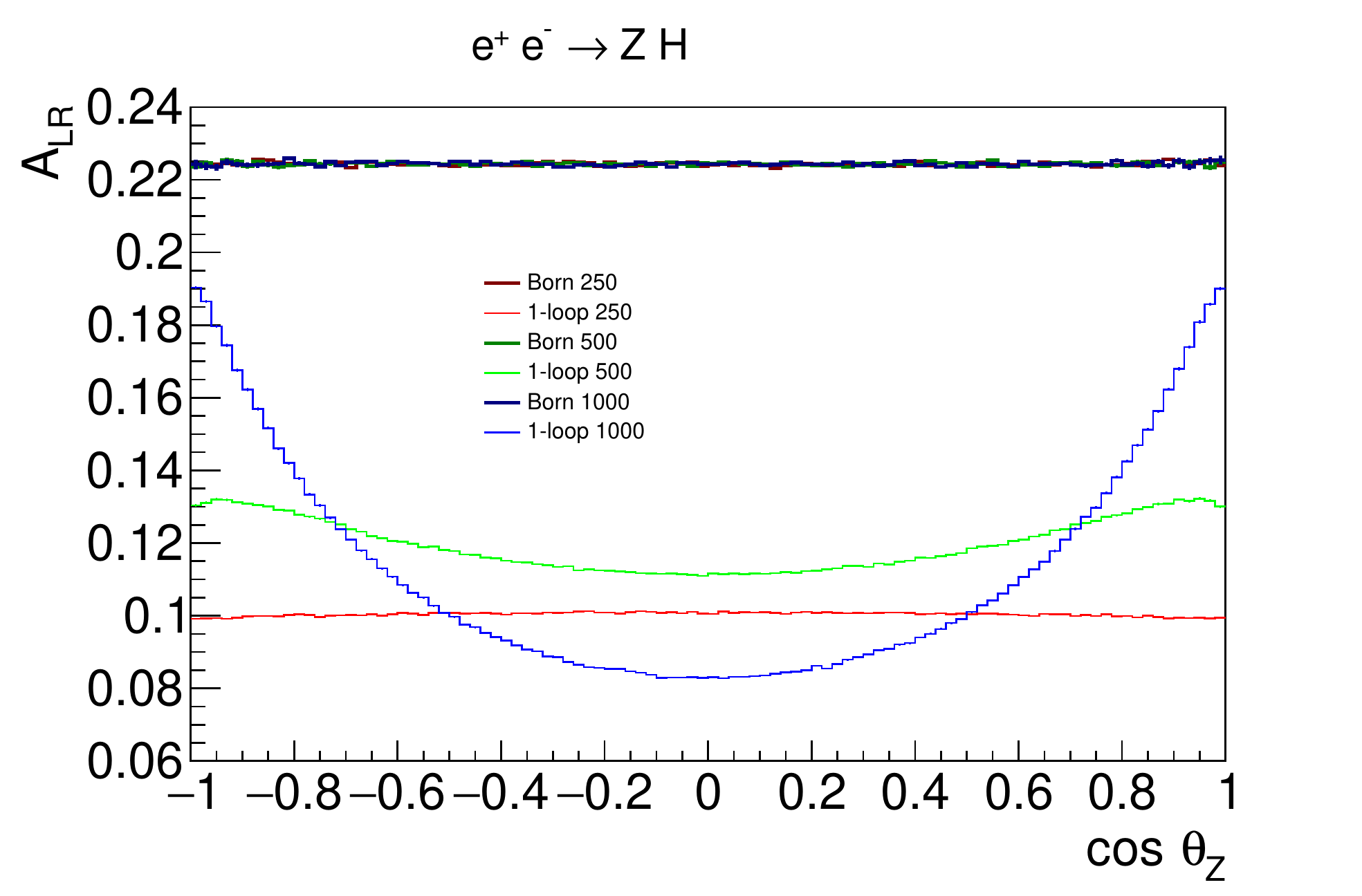}
\caption{Distributions in $\cos\vartheta_{Z}$ of the left-right asymmetry $A_{LR}$ for three different energies
$\sqrt{s} = 250, 500, 1000$ GeV obtained with {\tt SANC}.}
\label{ALR}
\end{center}
\end{figure*}

Figure~\ref{ALR} shows the distributions of left-right asymmetry $A_{LR}$ for three different energies
$\sqrt{s} = 250, 500, 1000$ GeV, where $A_{LR}$ is defined as
\bqa
A_{LR} = \frac{\sigma_{LR}-\sigma_{RL}}{\sigma_{LR}+\sigma_{RL}}.
\eqa

At Born level the $A_{LR}$ is constant:
\bqa
A_{LR}^{Born} = \frac{-3M^4_{\sss{Z}}+4\mzs \mws}{5M^4_{\sss{Z}}-12\mzs \mws+8M^4_{\sss{W}}} = 0.2243.
\eqa

\section{Conclusion}

In the paper we investigate the process $e^+e^- \to ZH$ at one-loop level with
longitudinal  polarizations of the positron and electron beams.

HA approach to the calculation of all components of the cross section: Born, virtual, soft part and hard Bremsstrahlung
makes it easy to take into account any polarization of the beams.

Table 1-3 summarizes the estimation of the  the correction value  $\delta$ in percent
for the set (0,0;-0.8,0;-0.8,-0.6;-0.8,0.6) of longitudinal  polarizations $P_{e^+}$ and $P_{e^-}$
of the positron and electron beams, respectively, and for the energies: $250, 500, 1000$ GeV.
Estimation of correction  $\delta$ amounts  significant  value: 6-20 $\%$ for our set of the
polarization value.

The asymmetry analysis shows a significant increase in $A_{LR}$
at high angles with the increasing energy (from $250$ GeV to $1000$ GeV).


\section{Acknowledgement}

Results were obtained within the framework of state's task
N 3.9696.2017/8.9 from Ministery of Education and Science of Russia.

\bibliographystyle{utphys_spires}
\addcontentsline{toc}{section}{\refname}\bibliography{ZH_v3}

\end{document}